# Input graph: the hidden geometry in controlling complex networks

Xizhe Zhang [1], Tianyang Lv [2,3], Yuanyuan Pu[1]

[1] （School of Computer Science and Engineering, Northeastern University, Shenyang110819, China）

[2] （College of Computer Science and Technology, Harbin Engineering University, Harbin 150001, China)

[3] （IT Center, National Audit Office, Beijing 100830, China)

## Abstract

The ability to control a complex network towards a desired behavior relies on our understanding of the complex nature of these social and technological networks. The existence of numerous control schemes in a network promotes us to wonder: what is the underlying relationship of all possible input nodes? Here we introduce input graph, a simple geometry that reveals the complex relationship between all control schemes and input nodes. We prove that the node adjacent to an input node in the input graph will appear in another control scheme, and the connected nodes in input graph have the same type in control, which they are either all possible input nodes or not. Furthermore, we find that the giant components emerge in the input graphs of many real networks, which provides a clear topological explanation of bifurcation phenomenon emerging in dense networks and promotes us to design an efficient method to alter the node type in control. The findings provide an insight into control principles of complex networks and offer a general mechanism to design a suitable control scheme for different purposes.

## Introduction

Controlling complex networked systems is a fundamental challenge in natural, social sciences and engineered systems. A networked system is controllable if its state can be controlled from any initial state to a desired accessible state[1-2] by inputting external signals from a few suitable selected nodes, which are called input nodes[3-6]. Existing works[3] provide an efficient method based on maximum matching to find a **M**inimum **I**nput nodes **S**et (abbreviated *MIS*) used to fully control a network.

However, these works have primarily focused on analyzing single *MIS*[4-8], while the underlying control relationships of nodes and *MIS*s remain elusive. Owing to the structural complexity of a network, its *MIS*s are typically not unique and the number of *MIS*s are exponential to the size of the network[9-10]. The enumeration of all possible *MIS*s is a #P problem[11] which requires high computational costs. A few works analyzed the node types in control[12-13] and control capacities[10] of input nodes. Moreover, although any of its *MIS*s are capable of fully controlling the network, they may composed of nodes with different topological properties, such as high-degree nodes[14]. The existence of physical constraints and limitations[15] may also affect the choice of a suitable *MIS*. For example, when controlling an inter-bank market[16-17,] one may need certain specific input nodes to ensure that a *MIS* can be manipulated by a given organization;

when controlling a protein interaction network[18], some proteins cannot be used as input nodes because of technique limitation.

Given the existence of numerous *MIS*s in a network, a node can be classified based on its participation in *MIS*s[12]: 1. possible input node, which appear in at least one *MIS*; 2. redundant node, which never appear in any *MIS*. Previous works[12] found that the dense networks exhibit a surprising bifurcation phenomenon, in which the majority of nodes are either redundant nodes or possible driver nodes. However, the origin of bifurcation phenomenon and the method of altering the type of nodes are still unknown.

Besides many approaches on controllability analysis of complex networks, the following questions are critical yet remain unknown: (*i*). what is the relationship between many available *MIS*s of a network? (*ii*). what topological structure determines whether a node is a possible input node? (*iii*). how to design suitable *MIS* with the desired nodes?

Here, we present input graph, a simple geometry but capable of revealing the complex correlation of all *MIS*s and nodes in control. The input graph is constructed by replacing the original edges with new edges reflecting control correlations of nodes. We prove that the node adjacent to an input node in input graph will appear in an *MIS*, and the nodes of the same connected component in input graph have the same control type, thus they are either all participate in control or not. Therefore, the emergence of giant connected component in input graph provides a clear topological explanation of the bifurcation phenomenon[12] in dense networks, and the complex control correlation of nodes of original network can be reduced into a few simple connected components of input graph. Furthermore, we present an efficient method to precisely manipulate the types of any node in control based on its connectivity of input graph. We believe that input graph is important because it (*i*) presents a framework that reveals the inherent correlation of *MIS*s and nodes in control and (*ii*) enables the design and manipulation of a suitable *MIS* of a network under constraints. Ultimately, this will promote the application of network control in real networked systems.

## Results

**Control adjacency and input graph**

The dynamics of a linear time-invariant network $G(V,E)$ is described by:

$$\frac{dx(t)}{dt} = Ax(t) + Bu(t) \qquad (1)$$

where the state vector $\boldsymbol{x}(t)=(x_1(t), \ldots, x_N(t))^T$ denotes the value of $N$ nodes in the network at time $t$, $A$ is the transpose of the adjacency matrix of the network, $B$ is the input matrix that defines how control signals are inputted to the network, and $\boldsymbol{u}(t)=(u_1(t), \ldots, u_H(t))^T$ represents the $H$ input signals at time $t$.

To analyze relationship of all nodes in control, we first define the control adjacency of nodes pair: for a network *G* and any maximum matching *M*, node *a* is said to be control adjacent to node *b* if there exists a node *c* connecting *a* and *b* with an unmatched edge $e_{ca}$ and a matched edge $e_{cb}$. Control adjacency reveals an important property about control correlation of nodes: a node that is control adjacent to an input node must appear in another *MIS*. For example, in Fig.1A, input node *a* is control adjacent to node *b*, and the two nodes alternately appear in *MIS* {*a*, *c*} and *MIS* {*b*, *c*}. We prove that this property is satisfied by any network, which is called the **Exchange Theorem**, that is: For any *MIS D* of a network *G*, if an input node *n*∈*D* is control adjacent to another node *m*, then *D'*=*D*\{*n*}∪{*m*} is also an *MIS* of *G* (see Supplementary Information). This means that a new *MIS D'* can be obtained from *MIS D* by exchanging a node of *D* with its control adjacent neighbor.

Then, we define the input graph $G_D(V, E_D)$ based on the control adjacency between nodes, where *V* is the node set, $E_D$ is the edge set and $e_{ij} \in E_D$ if node *i* is control adjacent to node *j* (Fig. 1C). Apparently, the input graph reveals all control relationships of nodes.

The input graph has several potential applications in analyzing controllability of complex networks. We find that the degree of a node in input graph reflects an important property about its substitutability in control. Based on the exchange theorem, for each edge of an input node in input graph, we can find a substitute node and obtain a new *MIS* with only one node replaced. It has important practical value. For example, when an input node of a *MIS* is no longer available due to physical constraint or attack, we can immediate find a new one by replacing the node with one of its neighbor in input graph. The above method makes the minimum change to the original control scheme, which is only one edge. Note that the computational complexity of the above process is only *O*(1), which yields a significantly improved method to obtain a new *MIS* in comparison with the state of the art approach[10].

**Connected components of input graph**

Next we focus on analyzing the connectivity of input graph. Similar to the concept of the path and reachable set in graph theory, we define control path *p* as the node sequence where neighbor nodes are control adjacent. The control-reachable set *C*(*n*) of node *n* is defined as the set of all nodes that are reachable from node *n* through any control path (Fig. 1B). Based on the above definition, we prove the following **Adjacency Corollary**: 1. For any *MIS D* and an input node *n*∈*D*, all nodes of *C*(*n*) must be possible input nodes; 2. For any *MIS D*, if node *m* did not belong to any control-reachable set of the input nodes of *D*, then *m* must be a redundant node and never appears in any *MIS* (see Supplementary

Information). Therefore, it is easy to conclude that a node is a possible input node if it can be control reachable from an input node of any MIS.

The adjacency corollary show that the control-reachable sets of possible input nodes and redundant nodes will never intersect. Thus, there are only two types of connected components in the input graph: 1. Input Component (*IC*), which contains at least one input node; and 2. Matched Component (*MC*), which contains no input node. We call the two type connected components as control components. Apparently, all nodes of *IC* are possible input nodes and all those of *MC* are redundant nodes.

We analyze the control components of some real networks, and find that the complex control relationships of these networks can be reduced into a few control components of input graphs, i.e., *little rock* (Fig.2A) and *political blog* networks (Fig.2B). Furthermore, we find that many real networks have a giant control component in their input graph (Table.1 and Fig.S7), suggesting that the majority of nodes are tightly connected by control adjacency and have the same type in control. The giant control component can either be a giant *IC*, or a giant *MC*.

To further understand the origin of giant control component, we analyze the size of the largest control component of synthetic networks. We found that the size of the largest control component increases with the average degree of a network (Fig.2C), whereas the number of control components decreases monotonically (Fig.2D). Therefore, there exists only one giant control component in dense networks (Fig.2E). The type of a node in control is determined by the type of control component to which it belongs, which depends on whether the control component contains an input node. If the giant control component contain at least one input nodes, most of its nodes will be possible input nodes; and if the giant control component contains no input node, most of its nodes will be redundant nodes. Thus, we can observe the bifurcation phenomenon[12] (Fig.2F) that emerges in dense networks. Therefore, the formation of the giant control component in input graph provides a clear explanation for the origin of the bifurcation phenomenon emergent in dense networks.

**Altering the type of nodes in control**

Owing to the economical or physical constraints exist in many actual control scenarios, we may need some specified nodes as input nodes. If the node is a possible input node, we can easily find a *MIS* which contain the node. However, if the node is a redundant node, we must alter the structure of the network and turn the node into a possible input node.

Since the nodes of the same control component have same control type, we only need to alter the type of control component in which the target node lies. This problem can be solved by adding new edges to the network. For example, if we link several input nodes to an *MC*, the nodes in the *MC* will be turned into possible input nodes and the *MC* will be turned into an *IC*. Additionally, if we match all input nodes of an *IC*, it will be turned into an *MC* and all nodes in it will be redundant nodes (Fig.3A-3B, 3E).

Therefore, we present an algorithm to alter the type of the control component (see Method). To quantify the efficiency of the algorithm, we investigate the number of added edges $p$ in both *ER-random* and *scale-free* networks. The results (Fig.3C-3D) showed that $p$ significantly decreases with the average degree $<k>$, and the proportion of changed possible input nodes $\Delta n_D$ increases monotonically, which indicates that it is easier to alter the control component of a denser network.

Surprisingly, the giant control component of a few networks can be changed by adding only one edge (Fig.3E-3F). For example, the control type of most nodes of some real networks (e.g. *Facebook* and *Amazon* networks shown in Table.1) can be altered by only one added edge. All these networks have a special giant *MC* in their corresponding input graphs, and the nodes of the *MC* was not linked by any unsaturated node (node without a matched out-edge) in the original network. We call it as a saturated matched component (*SMC*). Therefore, if we link an input node to an *SMC*, most nodes of the *SMC* will be control reachable by the input node and be turned into possible input nodes. However, when an *MC* is linked by one or more unsaturated nodes in original network, which we call it as an unsaturated matched component (*UMC*), we need to match all the unsaturated nodes to change its type. The result show the cost of altering an *IC* to a *UMC* (Fig.3C) is similar to that of altering a *UMC* to an *IC* (Fig.3D).

Furthermore, we find that the size of *MIS* significantly decreases after altering the type of the giant control component (Fig.S8), which means that the method can also be used to optimize the controllability of complex networks[19-20].

## Discussion

In summary, we developed the input graph, a fundamental structure that reveals the control relationship of nodes and *MIS*s. Our key finding, that the control adjacent nodes have the same type in control, allows us to reveal the inherent control correlation of nodes, and offers a general mechanism to manipulate the control type of nodes or design a suitable control scheme. Furthermore, networks with a giant control component display a surprising type transition phenomenon in response to well-chosen structural perturbations, which is ubiquitous in dense networks across multiple disciplines.

The input graph presented here is a starting point for deeply investigating the control properties of networks. It paves the way to analyze the properties of all *MIS*s of a network, such as enumerate all *MIS*s[14], estimate the node capacity in *MIS*s[10] or find the optimum *MIS* under different control constraints or with specific node property[21]. Furthermore, the structural properties of the input graph, such as the node's degree and connected component also reveal several important topics on controllability of a network. We believe the other structural properties such as average distance and diameter of input graph, also worth deep investigation for multiple disciplines, such as brain network[22], protein interaction[18], and et.al.

However, besides the input nodes in a *MIS*, there may exist other nodes which also need to be inputted the control signals. These nodes formed a cycle[4] in the network and cannot accessible from any input nodes in the current *MIS*. Therefore, to fully control the network, a control signal need to be inputted to any node of the cycle, and the signal can be shared with any input node[4]. Furthermore, there may exist more than one input nodes within the same input component. Any of these input nodes can be exchanged with its control reachable node based on exchange theorem. However, if two or more input nodes share same neighbor node in the input graph, they would not be substituted at the same time.

Furthermore, we design an algorithm to alter the control type of most nodes of a network with small structural perturbations, which is the first attempt to convert the control mode[12-13] of a network as far as we know. Many real networks, especially biological network, are incomplete and may have many missing edges. It means that if some new edges are discovered, it may alter the control type of existing nodes dramatically. However, these newly discovered edges will never weaken the performance of our algorithm, because they will only increase the size of the giant connected component of input graph.

Overall, these findings will improve our understanding of the control principles of complex networks and may be useful in controlling various real complex systems, such as drug designs[23-25], financial markets[16,26] and biological networks[18,22].

## Methods

**Construct input graph**

To build an input graph $G_D(V, E_D)$ of the directed network $G(V, E)$, we need find all control adjacent edges $E_D$ between nodes. Based on the adjacency corollary, there are no control adjacent relationship between any possible input nodes and redundant nodes. Therefore, the set of edges $E_D$ of input graph are composed by two subsets: the set of edges $E_{Di}$ between all possible input nodes and the set of edges $E_{Dr}$ between all redundant nodes.

The edges set $E_{Di}$ and all possible input nodes $V_{PD}$ can be found as follows:

1. Find a maximum matching *M* and the corresponding *MIS D*; let the candidate set of all possible input nodes $V_{PD}=D$;

2. Select a node *n* of *D*, let $D=D-\{n\}$; for all in-edges of node *n*, find the corresponding control adjacent neighbors $\{c_1,c_2,\ldots,c_i\}$;

3. Let $D=D + \{c_1,c_2,\ldots,c_i\}$, $V_{PD}=V_{PD} + \{c_1,c_2,\ldots,c_i\}$ and $E_{Di}= E_{Di} +\{e(n, c_1),\ldots,e(n, c_i)\}$;

4. Repeat step 2 and 3 until *D* is empty.

The edges set $E_{Dr}$ between all redundant nodes can be found as follows:

1. Let $V_{temp}= V- V_{PD}$;
2. Select a node $n$ of $V_{temp}$, let $V_{temp} \leftarrow V_{temp} -\{n\}$;
3. Let the matched in-edge of $n$ be $e(m,n)$, find all out-edge $\{e(m, c_1),\ldots,e(m, c_i) \}$ of node $m$; let $E_{Dr}= E_{Dr} +\{e(c_1,n),\ldots,e(c_i, n)\}$ and $V_{temp} \leftarrow V_{temp} +\{c_1,c_2,\ldots,c_i \}$;
4. Repeat step 2 and 3 until $V_{temp}$ is empty.

Next we analyze the computational complexity of above method. Let $N$ is the number of the nodes and $L$ is the number of the edges of the directed network $G(V, E)$. First, the complexity for finding a maximum matching is $O(N^{0.5}L)$[9]. Second, each node requires a breadth first search (*BFS*) process to finding its control reachable set, which computational complexity is $O(L)$. For the worst case, we need to find the control researchable set for all nodes and the complexity is $O(NL)$. Therefore, the total computational complexity to building an input graph is O(*NL*).

**Altering an *IC* to a *MC***

For a network $G(V,E)$, let $B(V_{out}, V_{in}, E)$ be the corresponding bipartite graph and $IC_{alter}$ be the target input component. The basic idea of altering an *IC* to a *MC* is to match all input nodes of the *IC* by adding edges. The following are the detailed steps:

1. Find all unmatched nodes $S$ corresponding to current maximum matching. Let $S_1=S \cap V_{out}$, $S_2= S \cap V_{in} \cap IC_{alter}$;

2. Select a node $n \in S_2$ and a node $m \in S_1$; add an edge $e_{mn}$ to $G$; remove nodes $n$ and $m$ from $S_2$ and $S_1$, respectively;

3. Repeat step 2 until $S_2$ is empty.

The correctness of method of above method is list as follows. First, we prove that $|S_1| \geq |S_2|$. Apparently, Based on the definition of $B(V_{out}, V_{in}, E)$, $|V_{in}|=|V_{out}|$. Because every edge of the maximum matching starts with a node of $V_{out}$ and ends with a node of $V_{in}$, then $|S_1|=|S \cap V_{out}|=|S \cap V_{in}| \geq |S \cap V_{in} \cap IC_{alter}|=|S_2|$, which means that for any node of $S_2$, we can find a corresponding node in $S_1$. When we add an edge $e_{mn}$ to $G$ in step 3, the matching $M'=M+e_{mn}$ must be a maximum matching of $G'=G+ e_{mn}$ because $n$ and $m$ are not matched by $M$. Therefore, nodes $n$ and $m$ are matched by $M'$. When we match all input nodes of the $IC_{alter}$, the $IC_{alter}$ will be turned into an *SMC*.

**Altering a *MC* to an *IC***

For a network $G(V,E)$ with a *MC*, let $B(V_{out}, V_{in}, E)$ be the corresponding bipartite graph. The basic idea of altering a *MC* to an *IC* is to link several input nodes to the *MC* and the *MC* will be turned into an *IC*. For a network with an *UMC*, if we directly link an input node to the *UMC*, the input node will be matched with an unsaturated node of the *UMC*. Therefore, we need first alter the *UMC* to an *SMC* and then turn the *SMC* into an *IC*. The method of altering an *UMC* to an *SMC* is similar to altering an *IC* to an *SMC*, which is as follows:

1. Find all unmatched nodes $S$ corresponding to current maximum matching. Let $S_1 = S \cap V_{in}$, $S_2 = S \cap V_{out} \cap UMC_{alter}$;

2. Select a node $n \in S_2$ and a node $m \in S_1$; add an edge $e_{nm}$ to $G$; remove nodes $n$ and $m$ from $S_2$ and $S_1$, respectively;

3. Repeat step 2 until $S_2$ is empty.

For a network with a giant *SMC*, we only need to link an input node to the *SMC*, and the part which is control reachable by the input node will be turned into an *IC*. Therefore, in order to maximize the size of result *IC*, we need to find the most "influential" node of the *SMC* based on the size of its control-reachable set. The algorithm is listed as follows:

1. Let $SMC_{alter}$ be the target saturated matched component; compute the size of control-reachable set of nodes in $SMC_{alter}$, let the node with maximum size be $n$;

2. Let $e_{mn}$ be the matched edge pointing to node $n$; select an input node $d$, add edge $e_{md}$ to $G$.

After adding the edge $e_{md}$, the nodes of control-reachable set of $n$ will be turned into possible input nodes because the input node $d$ is control adjacent to node $n$. If we want to convert all nodes of an *SMC*, we need to find the minimal input nodes set that can reach all node of the *SMC*. The problem can be solved by a simple greedy algorithm.

**Acknowledgments**

We thank Guangyan Zhang and Bin Zhang for discussions. This work was supported by the Fundamental Research Funds for the Central Universities of China under grand number N140404011, the Natural Science Foundation of China under grant number 60903009, 71272216, 91546110, the Special Program for Applied Research on Super Computation of the NSFC-Guangdong Joint Fund (the second phase), and the plan Project for Youth Scholar Backbone of General Colleges and Universities of Heilongjiang under the grant number 1253G017.

**Author Contributions**

X.-Z.Z. was the lead writer of the manuscript. X.-Z.Z. designed research, proved the theorems, analyzed the data and wrote the paper. T.-Y. L. wrote the paper and participated in designing the research. Y.-Y.P. performed the experiments. All authors reviewed the manuscript.

**Additional Information**

The authors declare no competing financial interests. Correspondence and requests for materials should be addressed to X.-Z.Z. (Email: zhangxizhe@mail.neu.edu.cn) or T.-Y.L. (Email: raynor1979@163.com)

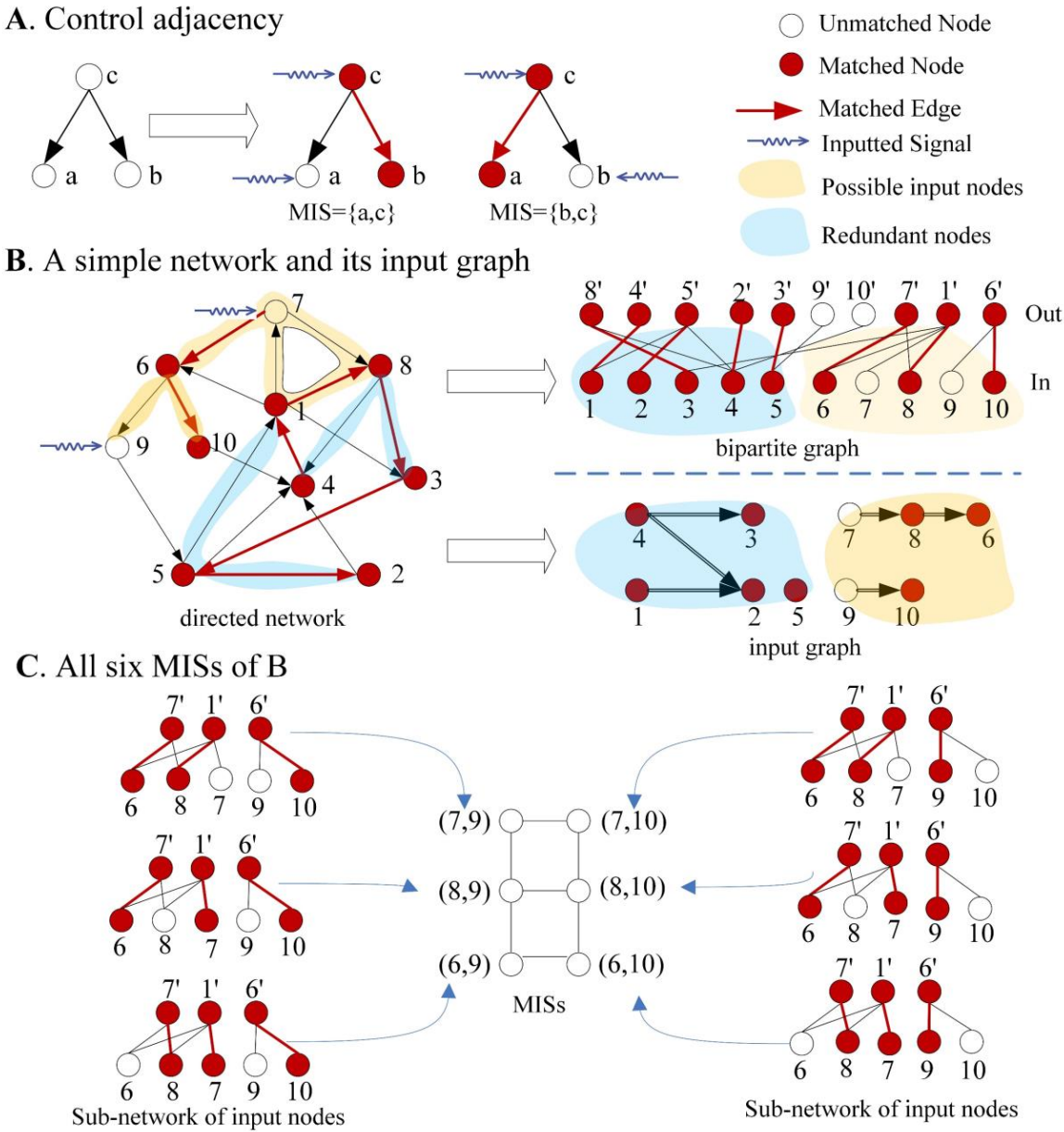

**Fig. 1.** Control adjacency and input graph. (**A**) A simple network with dilation. For a maximum matching {$e_{cb}$}, input nodes $a$ and matched node $b$ are control adjacent because they are connected by node $c$ with an unmatched edge $e_{ca}$ and a matched edge $e_{cb}$, which makes them alternately appear in two *MIS*s, {$a,c$} and {$b,c$}; (**B**) Sample directed network (left) and its bipartite graph (right up) and input graph (right down). The bipartite graph are constructed by split nodes of directed network into two separated nodes set in and out, which make a clear representation of control structure. The input graph are built based on control adjacency relationship. The input components (shaded in yellow) contain all possible input nodes and the matched components (shaded in blue) contain all redundant nodes. (**C**) All six *MIS*s of the network shown in (B), and the edges are their control adjacent relationship. If an input node of a *MIS* is no longer useable, we can immediate find its substitute node based on input graph. Note that the difference of adjacent *MIS*s are one matched edge and input node, which is the minimal change to original control structure.

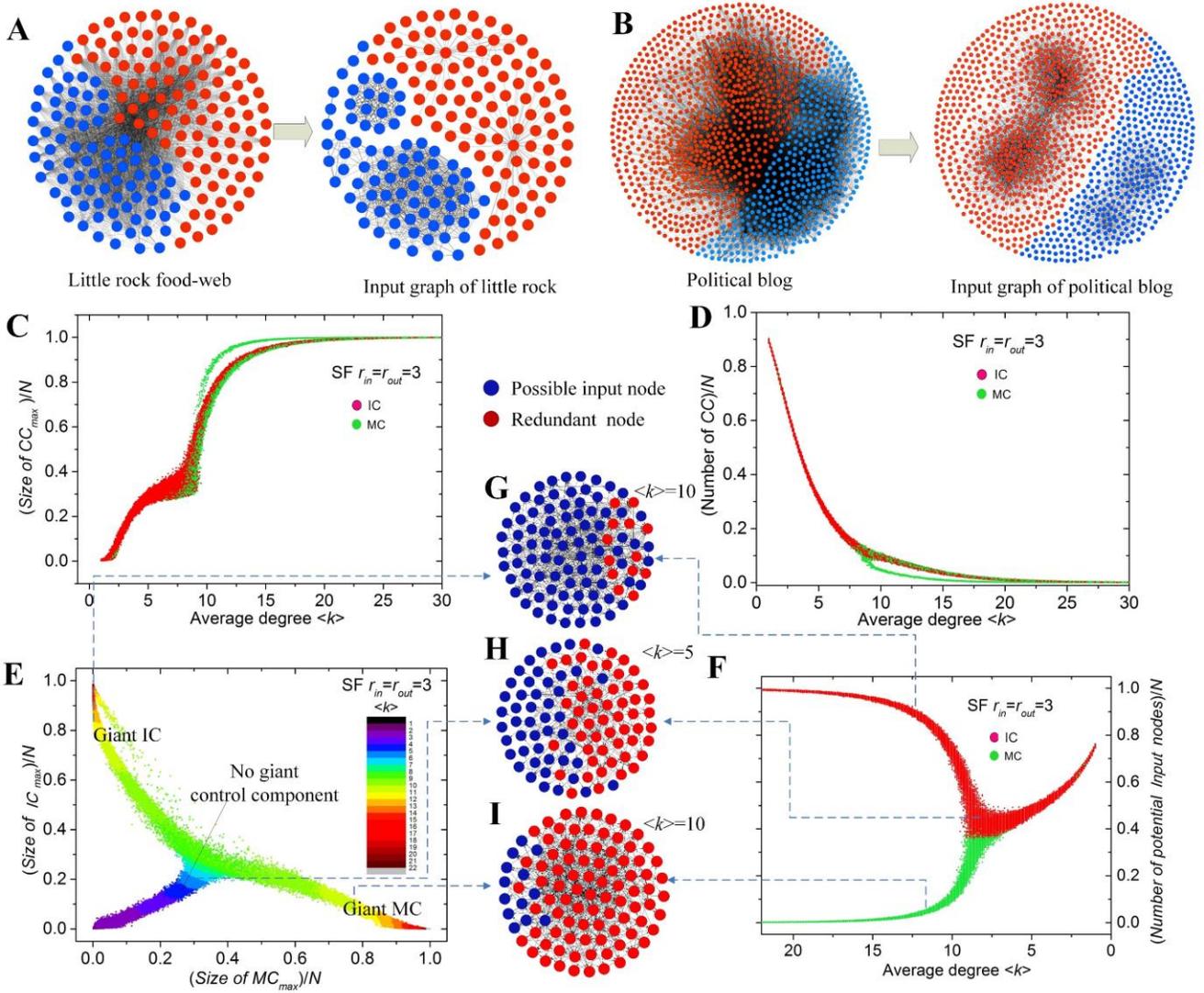

**Fig.2.** Control components of real and synthetic networks. (A-B) Real networks and its corresponding control adjacency network. The complex structure of (A) *Little rock food-web* and (B) *Political blog* are reduced into several simple connected components of input graph. (C) The size of the largest control component $CC_{max}$ versus the average degree $<k>$ in scale-free networks with degree exponents $r_{in}=r_{out}=3$, $N=10^4$ and (D) the number of control components ($CC$) decreases significantly with increasing $<k>$, which illustrates the emergence of a giant control component; (E) two types of giant control components; the input component (*IC*) and matched component (*MC*) cannot coexist in highly connected networks; (F) the emergence of a giant control component leads to the bifurcation phenomenon of possible input nodes in dense networks. The majority of nodes of a network with a giant *IC* are possible input nodes, whereas those of a network with a giant *MC* are redundant nodes; (G-H) three example networks with average degrees of $<k>=5$ and $<k>=10$; the type of their giant control component determines the type of the majority of nodes in control.

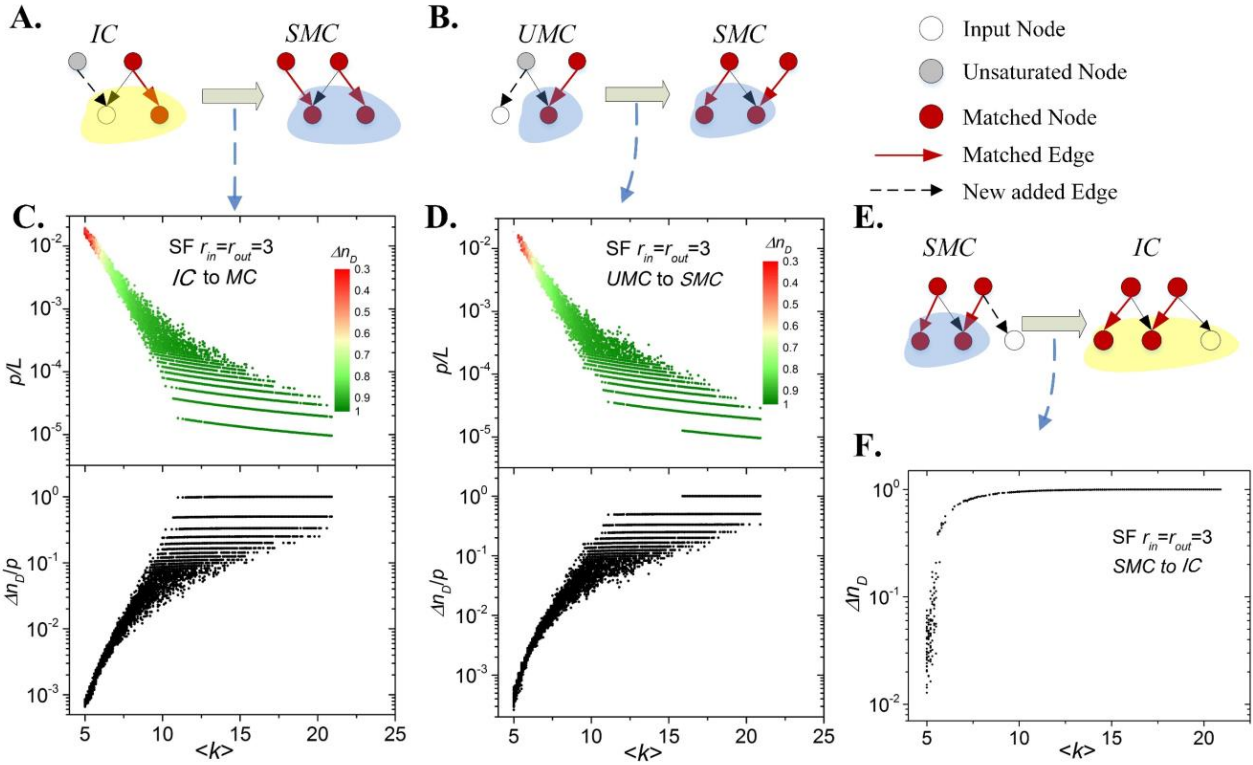

**Fig.3** | Type alteration of a giant control component in scale-free networks with degree exponents $r_{in}=r_{out}=3$, $N=10^4$. (**A**) Illustration of altering an input component (*IC*) to a saturated matched component (*SMC*). For each input node of an *IC*, we must add an edge pointing from an unsaturated node (node without matched out-edge) to the input node, and the *IC* will turn into an *SMC*; (**B**) Illustration of an alteration of an unsaturated matched component (*UMC*) to an *SMC*. For each unsaturated node of an *MC*, we add an edge from the unsaturated node to an input node, and the *UMC* will turn into an *SMC*; (**C-D**). The average degree versus percentage of added edges $p/L$ used to alter the control component for *IC* and *UMC* to *SMC*. For large $<k>$, the percentage of added edges significant decreases, and the changed possible input nodes of per edge $\Delta n_D/p$ increases rapidly, which indicates that it is easy to alter the giant control component type of dense networks; (**E**) Illustration of an alteration of an *SMC* to an *IC*. We need link an input nodes to *SMC* and it will turn into *IC*; (**F**) Average degree versus density of changed possible input nodes for *SMC* to *IC*. With only one added edge, the control type of most nodes changes in dense network.

**Table 1**. Characteristics of the real networks analyzed in the paper. For each network, we show its type, name, number of nodes (*N*) and edges (*L*), density of input nodes $n_{MIS}$, size and type of the largest control component $CC_{max}$, in which *I*, *U* and *S* denote *IC*, *UMC* and *SMC* respectively, the proportion of edges (*p*) that is added into the network to change the type of $CC_{max}$ and the density of changed possible input nodes ($\Delta n_D$) after adding edges.

| Type | Name | N | L | $<k>$ | $n_{MIS}$ | $CC_{max}$ | p | $\Delta n_D$ |
|---|---|---|---|---|---|---|---|---|
| Food Web | StMarks | 54 | 356 | 13.19 | 24.07% | 38.89%(U) | 3.37% | 62.96% |
| | Ythan | 135 | 601 | 8.90 | 51.11% | 85.19%(I) | 10.65% | 85.19% |
| | Mangrove | 97 | 1492 | 30.76 | 22.68% | 55.67%(I) | 1.41% | 55.67% |
| | Florida | 128 | 2106 | 32.91 | 23.44% | 91.41%(I) | 1.38% | 91.41% |
| | Silwood | 154 | 370 | 4.81 | 75.32% | 84.42%(I) | 29.19% | 84.42% |
| | Littlerock | 183 | 2494 | 27.26 | 54.10% | 34.43%(I) | 2.49% | 34.43% |
| Neuronal | C. elegans | 306 | 2345 | 15.33 | 18.95% | 68.63%(I) | 0.90% | 68.63% |
| Transcription | E.Coli | 423 | 578 | 2.73 | 72.81% | 12.53%(U) | 34.78% | 1.89% |
| | TRN-Yeast-1 | 4441 | 12873 | 5.80 | 96.46% | 99.21%(I) | 33.04% | 99.21% |

| | | | | | | | |
|---|---|---|---|---|---|---|---|
| | TRN-Yeast-2 | 688 | 1079 | 3.14 | 82.12% | 76.60%(I) | 41.61% | 76.60% |
| **Trust** | Prison inmate | 67 | 182 | 5.43 | 13.43% | 59.70%(U) | 2.75% | 74.63% |
| | Slashdot0902 | 82168 | 948464 | 23.09 | 4.55% | 91.23%(I) | 0.39% | 91.23% |
| | WikiVote | 7115 | 103689 | 29.15 | 66.56% | 32.12%(U) | 3.60% | 26.77% |
| **Electronic circuits** | s208a | 122 | 189 | 3.10 | 23.77% | 17.21%(I) | 5.82% | 17.21% |
| | s420a | 252 | 399 | 3.17 | 23.41% | 9.13%(I) | 3.26% | 9.13% |
| | s838a | 512 | 819 | 3.20 | 23.24% | 5.27%(I) | 2.08% | 5.27% |
| **Citation** | ArXiv-HepTh | 27770 | 352807 | 25.41 | 21.58% | 48.96%(U) | 0.91% | 51.89% |
| | SciMet | 3084 | 10416 | 6.75 | 37.48% | 52.14%(I) | 7.62% | 52.14% |
| | Kohonen | 4470 | 12731 | 5.70 | 47.29% | 60.25%(I) | 14.45% | 60.25% |
| **WWW** | Political blogs | 1224 | 16718 | 27.32 | 34.15% | 52.61%(I) | 1.46% | 52.61% |
| | NotreDame | 325729 | 1497134 | 9.19 | 67.71% | 53.90%(I) | 9.15% | 53.90% |
| | BerkStan | 685230 | 7600595 | 22.18 | 65.69% | 57.27%(I) | 2.78% | 57.27% |
| | Google | 875713 | 5105039 | 11.66 | 36.95% | 60.80%(I) | 3.97% | 60.80% |
| | Stanford | 281903 | 2312497 | 16.41 | 35.91% | 63.50%(I) | 2.83% | 63.50% |
| **Internet** | p2p-1 | 10876 | 39994 | 7.35 | 55.20% | 90.58%(I) | 14.88% | 90.58% |
| | p2p-2 | 8846 | 31839 | 7.20 | 57.78% | 90.55%(I) | 15.52% | 90.55% |
| | p2p-3 | 8717 | 31525 | 7.23 | 57.74% | 91.75%(I) | 15.58% | 91.75% |
| **Organizational** | Consulting | 46 | 879 | 38.22 | 4.35% | 97.83%(I) | 0.23% | 97.83% |
| **Social communication** | UClonline | 1899 | 20296 | 21.38 | 32.33% | 79.94%(I) | 2.84% | 79.94% |
| **Product co-purchasing** | Amazon0302 | 262111 | 1234877 | 9.42 | 3.23% | 49.55%(U) | 0.30% | 74.58% |
| | Amazon0312 | 400727 | 3200440 | 15.97 | 3.52% | 83.61%(S) | 0.00003% | 75.74% |
| | Amazon0505 | 410236 | 3356824 | 16.37 | 3.62% | 91.35%(I) | 0.44% | 91.35% |
| | Amazon0601 | 403394 | 3387388 | 16.79 | 2.04% | 75.90%(U) | 0.21% | 90.73% |
| **Social network** | twitter_combined | 81306 | 1768149 | 43.49 | 19.39% | 79.40%(I) | 0.88% | 79.40% |
| | Facebook_0 | 347 | 5038 | 29.04 | 5.48% | 86.46%(S) | 0.02% | 81.84% |
| | Facebook_107 | 1912 | 53498 | 55.96 | 45.92% | 54.08%(S) | 0.001% | 53.24% |
| | Facebook_348 | 572 | 6384 | 22.32 | 61.01% | 38.64%(S) | 0.02% | 38.29% |

# Input graph: the hidden geometry in controlling complex networks

Xizhe Zhang [1], Tianyang Lv [2,3], Yuanyuan Pu[1]

[1] (School of Computer Science and Engineering, Northeastern University, Shenyang110819, China)

[2] (College of Computer Science and Technology, Harbin Engineering University, Harbin 150001, China)

[3] (IT Center, National Audit Office, Beijing 100830, China)

**Supplementary Information:**

Materials and Methods

Figures S1-S8

Tables S1-S2

.

# Materials and Methods

## 1. Background

We consider a linear time-invariant system $G(A, B)$, whose states are determined by the following equations [1]:

$$\frac{dx(t)}{dt} = Ax(t) + Bu(t) \tag{1}$$

where the state $\mathbf{x}(t)=(x_1(t),\ldots, x_N(t))^T$ denotes the value of all nodes at time $t$; $A$ is the transpose of the adjacency matrix; $\boldsymbol{u}(t)=(u_1(t),\ldots,u_M(t))^T$ is the input signal; $B$ is the input matrix that defines how control signals are inputted into the system.

The above system $G(A, B)$ is considered to be controllable [2-3] if it can be driven from the initial state to any admissible final state, which can be determined by Kalman's controllability rank condition [3], that is, the network is controllable if and only if the following matrix has full rank:

$$C = (B, AB, A^2B,\ldots, A^{N-1}B) \tag{2}$$

In many real control scenarios, the edge weight of a network is often unknown or time-variant. To overcome this difficulty, Lin [4] introduced structural controllability, which considers a network where only the structure of the zero and nonzero elements is known. If a network is structural controllable, it remain structural controllable for almost all weight values. To fully control the network, we need to input external signals to the nodes of a network. Because one input signal can be connected to multiple nodes, we call those nodes which do not share input signals as input nodes. The input nodes are also called driver nodes [5-8]. The Minimum set of independent input nodes that is used to fully control a network is called **M**inimum **I**nput nodes **S**et (*MIS*) [5-8].

An *MIS* can be determined by the maximum matching of a network [9], the unmatched nodes corresponding to any maximum matching are input nodes. Liu et.al [5] found that the input nodes tends to avoid high degree nodes and the size of an *MIS* is mainly determined by the degree distribution of a network. Ruths [6] quantified the node composition of an *MIS* and found that most real networks form three well-defined clusters. Menichetti [7] found that the fraction of input nodes is primarily determined by low in-degree and low out-degree nodes.

However, the maximum matchings of a network are generally not unique, as are *MIS*s. Jia [8] classified the nodes based on their participation in all *MIS*s: 1. critical input node, which appear in all *MIS*s; 2. redundant node, which never appear in any *MIS*; 3. intermittent input node, which appears in one or more *MIS*s. They found that the density of intermittent input nodes exhibit a surprising bifurcation phenomenon in dense networks, in which the majority of nodes are either intermittent input nodes or redundant nodes. However, because enumerating all possible maximum matching is a #P problem [10], it is very difficult to analyze all *MIS*s of large-scale networks. Jia et.al [11] analyzed the bifurcation phenomenon based on the greedy leaf removal (GLR) procedure [12]. They found the types of the core determined the type of majority nodes in control [8]. However, we still lack knowledge regarding the correlation of all input nodes and the method to alter the type of nodes in control.

**2. Maximum matching**

In this section we will introduce some basic concepts and theorems about maximum matching. The following contents can be found in the textbook of graph theory [13].

A graph $B$ ($V_1$, $V_2$, $E$) is called a bipartite graph if there are no edge connect nodes of $V_1$ and $V_2$. A set of edges in $B$ ($V_1$, $V_2$, $E$) is called a matching $M$ if no two edges in $M$ have a node in common. A node $v_i$ is said to be matched by $M$ if there is an edge of $M$ linked to $v_i$; otherwise, $v_i$ is unmatched. A path $P$ is said to be $M$-alternating if the edges of $P$ are alternately in and not in $M$. An $M$-alternating path $P$ that begin and ends at the unmatched nodes is called an $M$-augmenting path. The maximum matching is a matching with the maximum number of edges of the graph.

For any matching $M$, if there exist an $M$-augmenting path $P$, we can obtain a larger matching $M'$ by the symmetric difference of $M$ and $E(P)$, that is, $M' = M\Delta E(P) = (M\backslash E(P)) \cup (E(P)\backslash M)$. Therefore, the maximum matching can be found by searching the $M$-augmenting paths of the network, which are proved by the following theorem:

**Berge Theorem** [14]: Let $M$ be a matching of bipartite graph $B$, $M$ is a maximum matching if and only if there exist no $M$-augmenting path in $B$ corresponding to $M$.

A direct inference of above theorem is: for a given maximum matching $M$, if $P$ is an $M$-alternating path which start or end by one unmatched node, the symmetric difference of $M$ and $E(P)$ will result a different maximum matching $M'$.

Next we will introduce the maximum matching of a directed network which used to analyze the structural controllability [9]. Consider a directed network $G(V,E)$, where $V(G)$ is the node set and $E(G)$ is the edge set. To analyze the structural controllability, we first define its corresponding bipartite graph $B(V^{out},V^{in},E)$ as the follows: $V^{out}$ is the set of nodes with out-edges in $V$, and $V^{in}$ is the set of nodes with in-edges in $V$. A directed link $e_{ij}$ corresponds to a connection between node $i$ of $V^{out}$ and node $j$ of $V^{in}$ in the bipartite graph. (see Figure S1 for a detailed example).

An *MIS* can be obtained by computing the maximum matching of $B(V^{out},V^{in},E)$, the unmatched nodes in $V^{in}$ are input nodes [9], which can be inputted external control signals. The unmatched nodes in $V^{out}$ are called unsaturated nodes. Because $B(V^{out},V^{in},E)$ is constructed based on the directed network $G(V,E)$, therefore, for directed network $G$, the input nodes have no matched in-edge and the unsaturated nodes have no matched out-edge.

**3. Theorem**

## 3.1 Exchange Theorem

**Definition 1:** Control adjacency: For a network $G$ and its maximum matching $M$, consider nodes $a$ and $b$, if there exist a node $c$, an unmatched edge $(c,a)$ and a matched edge $(c,b)$, we say that $a$ is control adjacent to $b$.

**Exchange Theorem:** For a network $G$ and one of its *MIS D*, if there exist a input node $a \in D$ with non-zero in-degree $k_{in}(a) \neq 0$, then another node $b$ which is control adjacent to node $a$ must exist, and $D'=D\backslash\{a\}\cup\{b\}$ is also an *MIS* of $G$.

**Proof:** Let $M$ be the maximum matching corresponding to $D$. Because $k_{in}(a) \neq 0$, let $e_1=(c,a)$ be an in-edge of node $a$. First, we prove that node $c$ has a matched out-edge. If $c$ has no matched out-edges, $c$ must be an unmatched node, which means that edge $e_1=(c,a)$ connects two unmatched nodes $c$ and $a$ ($a$ is a input node of $D$). Therefore, based on the definition of matching, $M \cup \{e_1\}$ is also a matching of G. This contradicts to the fact that $M$ is a maximum matching of $G$. Therefore, $c$ must have a matched out-edge, denoted as $e_2=(c,b)$. By definition 1, the node $a$ is control adjacent to node $b$.

The next step is show that $D'=D\backslash\{a\}\cup\{b\}$ is an *MIS* of $G$, that is to prove that $M'=M\backslash\{e_2\}\cup\{e_1\}$ is another maximum matching of $G$. For maximum matching $M$, we already know that node $a$ have no matched in-edge (because it is a input node), $e_1=(c,a)$ is a unmatched edge and $e_2=(c,b)$ is a matched edge. Therefore, let $e_1=(c,a)$ be the matched edges and $e_2=(c,b)$ be the unmatched edge, according to the definition of matching, $M'=M\backslash\{e_2\}\cup\{e_1\}$ is still a matching of $G$ (see Figure.S2A). Clearly, $M'$ is a maximum matching because it have same number of edges as maximum matching $M$. Therefore, the unmatched nodes $D'=D\backslash\{a\}\cup\{b\}$ that corresponds to $M'$ is an *MIS* of network $G$. The proof is complete.

**Inference 1:** A node appears in all *MIS*s if and only if it has no in-edge.

**Proof:** Sufficiency. Based on exchange theorem, if node $n$ appeared in all *MIS*s, there must not exist a node that is control adjacent to $n$, which means that the $k_{in}(n)=0$.

Necessity. Suppose that $k_{in}(n)=0$. Because no edge point to node $a$, it is unmatched for any maximum matchings. Therefore, node $n$ appears in all *MIS*s.

**Inference 2:** The number of control adjacent neighbors of input nodes $n$ equals its in-degree.

**Proof:** Based on exchange theorem, for each in-edge of node $n$, there must exist a control adjacent node. Because the matched edges do not share nodes, therefore, each in-edge of $n$ correspond to different control adjacent node. The proof is complete.

## 3.2 Input graph and control components

**Definition 2:** Input graph: For a network $G(V,E)$ and one of its maximum matching $M$, the input graph $G_D(V,E_D)$ are defined based on the control adjacent relationship of nodes set $V$, where $E_D$ is the edge set and $e_{ij} \in E_D$ if node $i$ is control adjacent to node $j$.

**Definition 3:** Control path: For an edge sequence $P(e_1,e_2,\ldots,e_k)$ of $G$, we say that $P$ is a control path if and only if for any two adjacent edges $e_i(x,a)$ and $e_{i+1}(x,b)$ in $P(e_1,e_2,…,e_k)$, node $a$ is control adjacent to $b$. We say that node $a$ is control reachable to node $b$ if there exist a control path connecting them.

**Definition 4:** Control-reachable set: All nodes that are control reachable from input node $a$ are called the control-reachable set of $a$, denoted as $C(a)$. Note that $a \in C(a)$. The control-reachable set of *MIS D* is the union of its input nodes' control-reachable set, that is, $C(D) = \bigcup_{n \in D} C(n)$. For a matched node $b$, the control-reachable set $C(b)$ is defined as the set of nodes that is control reachable to $b$.

For nodes $n$ and $m$, $C(n)$ and $C(m)$ are connected if $C(n) \cap C(m) \neq \emptyset$. The control-reachable set of nodes of a network may connect with each other, and form the following connected components:

**Definition 5:** Control component: For any maximum matching of network *G*, the maximal connected sets of the control-reachable set of *G* are called the control components of *G*.

The control component can be classified as the following: 1. an input component (*IC*), which contains at least one input node (node without any matched in-edge); 2. a matched component (*MC*), which does not contain any input node.

**Lemma 1:** A control component cannot both contain an input node and link by an unsaturated node of *G*.

**Proof:** For network *G* and maximum matching *M*, let *C* be a control component and contains an input node *n*. Let node *m* be an unsaturated node. According to the definition of unsaturated node and input node, nodes *n* have no matched in-edge and node *m* have no matched out-edge.

Suppose node *m* is connected to a node of *C*, which have two following cases (Figure.S2B):

1. Node *m* is connected to input node $n \in C$. Because *n* and *m* are both unmatched nodes in the corresponding bipartite graph of *G*, that means that $e(m,n) \notin M$. Therefore, $M'=M + e(m,n)$ is another matching of *G*. That contradict to the fact that *M* is a maximum matching because $|M'|>|M|$.

2. Node *m* is connected to another matched node $k \in C$. Based on the definition of control component, there must exist an input node *j* which control reachable to *k*. Let the control path from *j* to *k* be $p(n,k)$, clearly, $p(j,k)+e(m,k)$ is an *M*-augmenting path because the node *m* and node *j* are both unmatched in the corresponding bipartite graph of *G*. That means that *M* is not a maximum matching based on Berge theorem [12]. This leads to a contradiction. Therefore, a control component cannot both contain an input node and link by an unsaturated node.

### 3.3 Adjacency Corollary

Because the input graph is constructed based on the maximum matching of the network, different maximum matchings may result different input graphs. However, the types of nodes in control and the types of control components are remain same for any maximum matching, which is proved by the following theorem.

**Adjacency Corollary 1:** For any *MIS D* and input node $a \in D$, all nodes of $C(a)$ are possible input nodes.

**Proof:** Let the maximum matching related to *MIS D* be *M*. For any node b∈*C*(*a*), based on the definition of a control-reachable set, there must exist a path $P_{ab}$ which starts with unmatched node *a* and ends with matched node *b*. Based on the definition of section 2, $P_{ab}$ is an *M*-alternating path.

Let *M'* be the symmetric difference of *M* and $E(P_{ab})$, that is, $M' = M\Delta E(P_{ab})$. Based on the inference of Berge theorem mentioned in section 2, *M'* is another maximum matching because $P_{ab}$ is an *M*-alternating path. Clearly, node *b* is not matched by *M'* because it is matched by *M* (Figure.S2C). Therefore, node *b* is an input node corresponding to maximum matching M'. The proof is complete.

Before we prove the Adjacency Corollary 2, we give the following property [15] about the symmetric difference of two maximum matching:

**Property 1** [15]: For two different maximum matching *M* and *M'*, each connected component of the symmetric difference $M\Delta M' = (M\backslash M') \cup (M'\backslash M)$ is one of the following (see Figure.S3):

(1) An isolated node.

(2) An even cycle with edges alternatively in *M\M'* and *M'\M*, or

(3) A path whose edges are alternatively in *M\M'* and *M'\M*.

**Adjacency Corollary** 2: For any *MIS D*, if node *b*∉*C*(*D*), *b* must be a redundant node.

**Proof:** Let *M* be the maximum matching corresponding to *MIS D*. Based on the definition of control reachable set, node *b* is matched by *M* and has one matched in-edges $e_{nb}$∈*M* because *b*∉*C*(*D*). Suppose that *b* appears in another *MIS D'* and *M'* is the corresponding maximum matching. Because we try to prove node *b* is not a possible input node, therefore, we only consider the in-edge of node *b* in the following proof. Clearly, all in-edges of node *b* are not matched corresponding to *M'*.

Now we consider the symmetric difference $M\Delta M'$. Let *CP*(*b*) be the connected component of the symmetric difference $M\Delta M'$ which contained node *b*. Based on Property 1, the symmetric difference of two maximum matchings have three cases. Note that node *b* is matched by *M* and is not matched by *M'*. Therefore, *CP*(*b*) is not an isolated node, or an even cycle (node *4* and *5* in Figure.S4D), or a path (node *6* in Figure.S4D), because node *b* are matched by both *M* and *M'* in all three cases. Therefore, *CP*(*b*) can be only *C*1 (node 2 or 3) in Figure.S4D: path *P* with an even length that starts with node *b* and ends with another node *n* of $V^{in}$.

Next we show that the end node *n* of path *CP*(*b*) is an unmatched node corresponding to maximum matching *M*. Because path *CP*(*b*)∈$M\Delta M'$, the edges of path *CP*(*b*) alternately appears in *M* and *M'*. Because the length of *CP*(*b*) is even, and node *b* is matched by *M*, it is easy to conclude that node *n* is not matched by *M*.

Finally, we prove that node *n* is control reachable to node *b*. Because node *b* is matched by *M*, *n* is not matched by *M* and the edges of *CP*(*b*) alternately appears in and out *M*. Therefore, based on the definition of control reachable set, node *n* is control reachable to node *b* corresponding to maximum matching *M*, that is *b*∈*C*(*a*)∈*C*(*D*). This leads to a contradiction. Therefore, *b* never appears in any *MIS* and must be a redundant node.

Based on the adjacency corollaries 1 and 2, we obtain the following important inference：

**Inference 4**: The control-reachable set of any *MIS D* is the union of all *MIS*s of a network.

**Proof**: Based on the theorem 3 and 4, the proof is trivial.

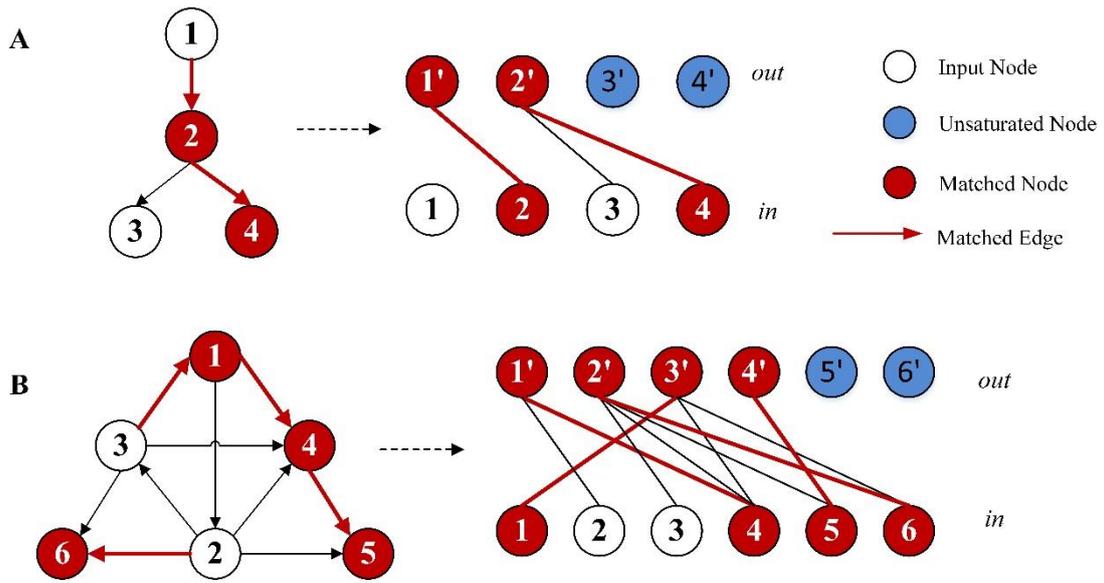

**Figure S1:** Two sample directed networks and their corresponding bipartite graphs. The red nodes and edges are matched nodes and matched edges of the maximum matchings. A node *n* of a sample network corresponds to two nodes of the bipartite graphs that belong to the *in* set and the *out* set of nodes respectively, recorded as *n* and *n'*. An edge $a_{n,m}$ of a sample network corresponds to an edge $e_{n',m}$ of the bipartite graph, where *n'* belongs to the *out* set and *m* belongs to the *in* set.

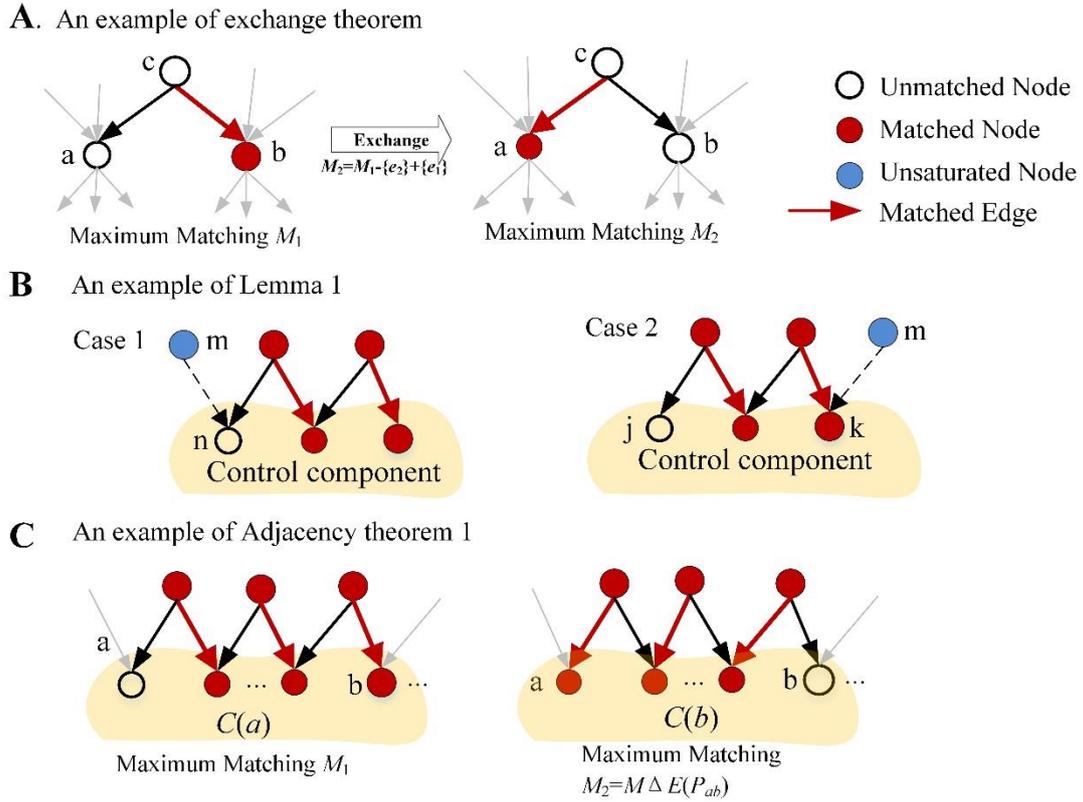

**Figure S2:** Some examples used in the proof of theorems. **A.** An example network and one of its maximum matching, node *a* is an input node and *b* is a matched node. If we exchange the matched edge $e_{cb}$ and the unmatched edge $e_{ca}$, we will get a new *MIS* in which node *b* is an input node and node *a* is a matched node; **B.** An example network used in the proof of Lemma 1. Case1: If unsaturated node *m* is connected to node *n*, edge $e_{mn}$ will be a matched edge because *m* and *n* are both unmatched by current maximum matching; Case 2: if node *m* connected to a matched node *j*, then path $P_{mj}$ will be a *M*-augmenting path because the node *m* and node *j* are both unmatched; **C.** An example network used in the proof of Adjacency Corollary 1. For a maximum matching $M_1$, if node *b* is control reachable by input node *a*, then $P_{ab}$ must be an *M*-alternating path. The symmetric difference of *M* and $E(P_{ab})$ $M_2$ is also a maximum matching, and the node b is not matched by $M_2$.

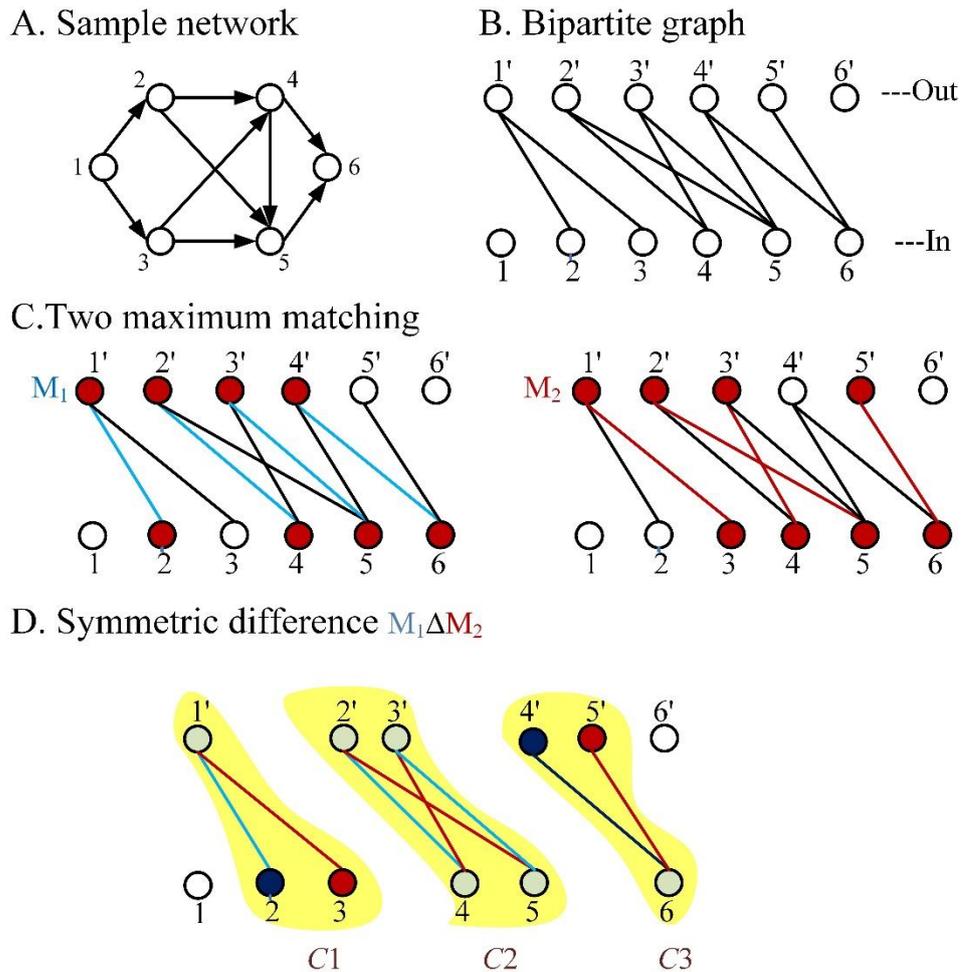

**Figure S3**: Symmetric difference of two maximum matchings of a sample network. (a) Sample network and its bipartite graph in (b); (c) two maximum matchings $M_1$ and $M_2$ of (b); (d) Symmetric difference of $M_1$ and $M_2$. There are three cases of its connected components: $C1$, $C2$ and $C3$. $C1$ and $C3$ are the paths whose edges are alternatively appear in $M_1\backslash M_2$ and $M_2\backslash M_1$. $C2$ is the even cycle with edges alternatively in $M_1\backslash M_2$ and $M_2\backslash M_1$. If $M_1$ and $M_2$ are maximum matchings corresponding to two different *MDS*s, the connected components of symmetric difference of $M_1$ and $M_2$ will only contain $C1$ type component.

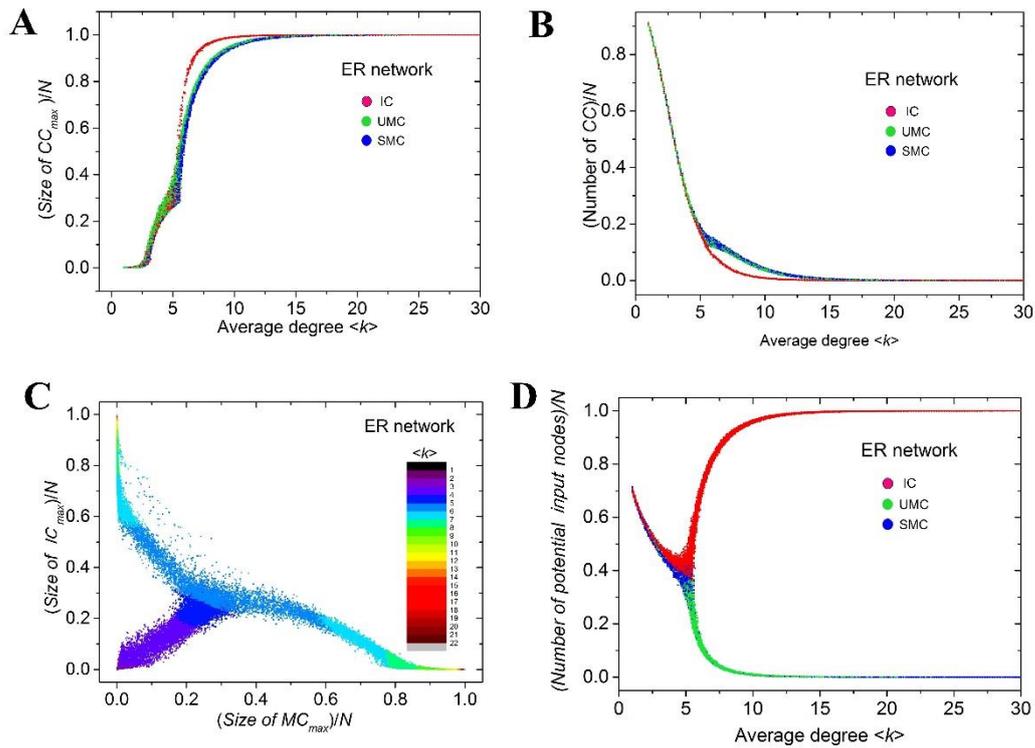

**Figure S4**: Control component of *ER* networks. The number of nodes $N=10^4$. (A). The size of the largest control component $CC_{max}$ versus the average degree $<k>$ in *ER* networks and (B) the number of control components (*CC*) decreases significantly with the increasing $<k>$, which illustrates the emergence of a giant control component; (C) two type of the giant control components, the input component (*IC*) and the matched component (*MC*) cannot coexist in the highly connected networks; (D) the emergence of the giant control component leads to the bifurcation phenomenon of possible input nodes in dense networks. The majority of nodes of a network with a giant *IC* are possible input nodes, whereas those of a network with a giant *MC* are redundant nodes.

## A. IC to SMC

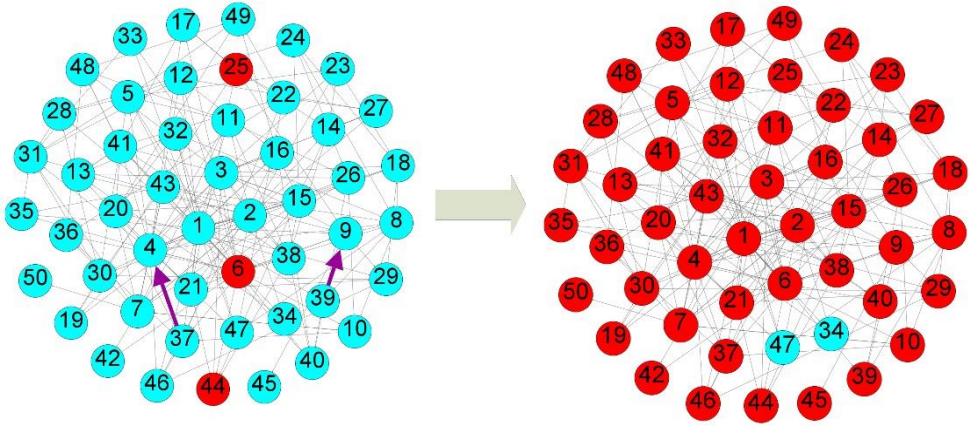

## B. UMC to SMC

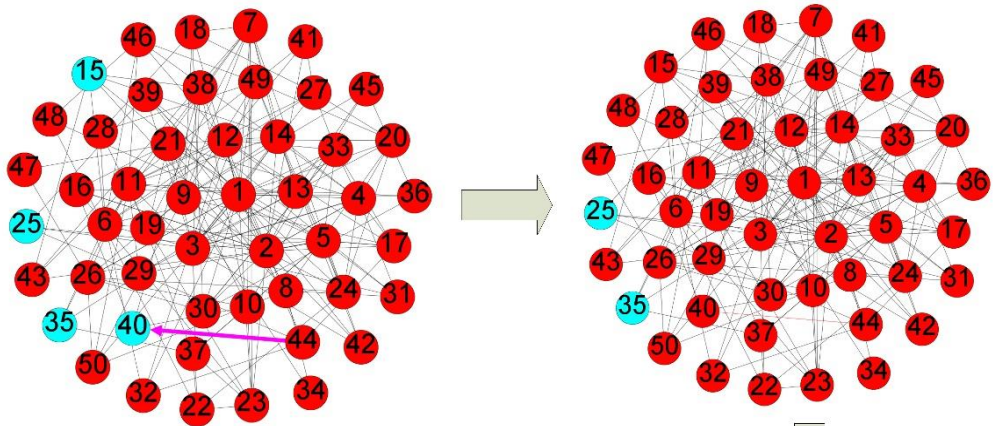

## C. SMC to IC

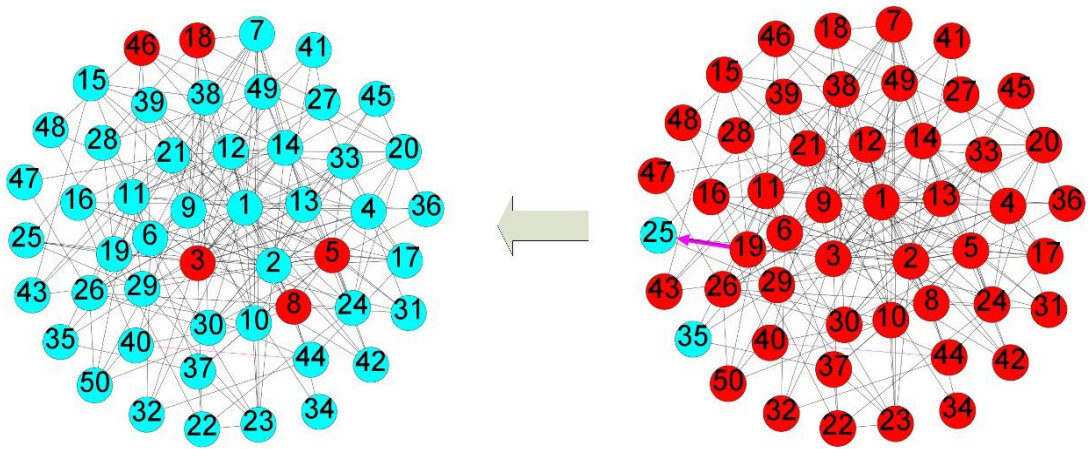

**Figure. S5** | Illustration of altering the type of the giant control component of sample networks. The number of nodes *N*=50. (**A**) Alternation of an *IC* to an *SMC* by adding two edges. After add the edges *e*(37,4) and *e*(39,9), most possible input nodes (blue nodes) are turned into redundant nodes (red nodes); (**B**) Alternation of an *UMC* to an *SMC* by adding one edge. After adding the edge *e*(44,40), the *UMC* is turned into an *SMC*, and (**C**) After adding the edge *e*(19,25), most redundant nodes (red nodes) are turned into possible input nodes (blue nodes), and the giant control component is turned into a giant *IC*.

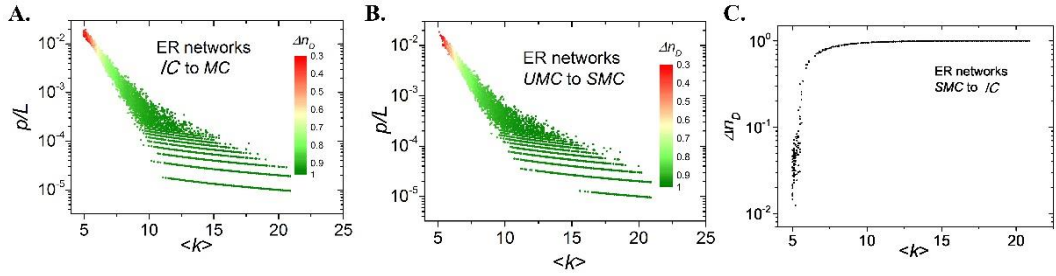

**Figure. S6** | Percentage of added edges *p/L* used to alter the giant control component versus average degree <*k*> of ER networks. The number of nodes $N=10^4$. (**A**) When one alter an *IC* to an *UMC*, the percentage of added edges significant decreases with increasing <*k*>, and the number of changed possible input nodes of each added edge $\Delta n_D/p$ increases rapidly; (**B**) When one alters an *UMC* to an *SMC*, the percentage of added edges used to alter the giant control component is similar to the case that one alters an *IC* to an *UMC*; (**F**) When one alters an *SMC* to an *IC*, the control type of most nodes in a dense network will be changed by adding only one edge.

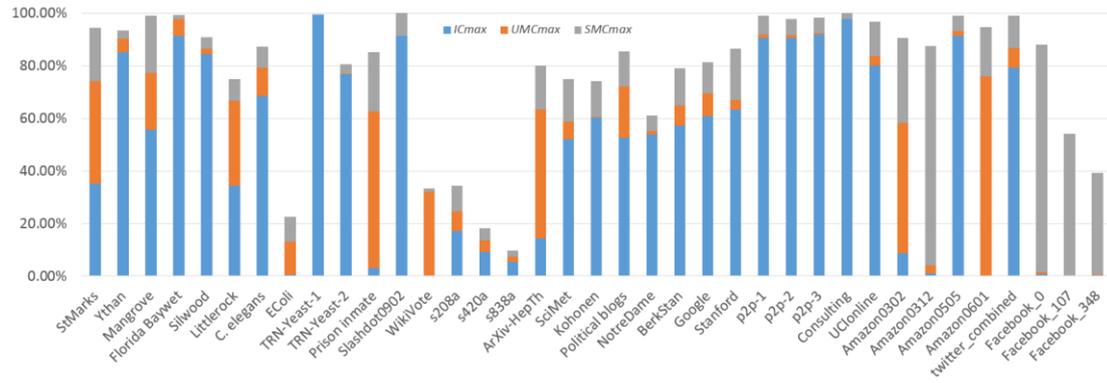

**Figure.S7** | Control component of real networks. The fractions of $IC_{max}$ (blue), $UMC_{max}$ (orange) and $SMC_{max}$ (grey) for the real networks named in Table 1. Note that we only counted the maximum control component of each type, thus the sum of fractions may be less than 100%.

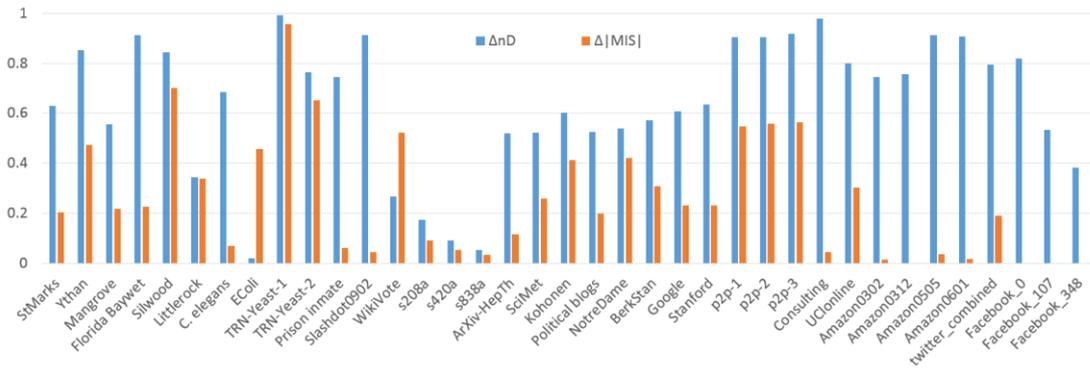

**Figure.S8** | The changed fraction of possible input nodes (blue) and the size of *MIS* (orange) after type transition of real networks shown in Table 1. The size of *MIS* significant decreases after type transition, which makes the networks easier to be controlled.

**Table S1.** Real networks analyzed in this paper. For each network, we show its type, name, number of nodes (*N*) and edges (*L*), and brief description.

| Type | Name | *N* | *L* | Description |
|---|---|---|---|---|
| Food Web | StMarks[16] | 54 | 356 | Food Web in St Marks national wildlife refuge. |
| | Ythan[17] | 135 | 601 | Food Web in Ythan Estuary. |
| | Mangrove[18] | 97 | 1492 | Food Web in Mangrove Estuary, Wet Season |
| | Florida[18] | 128 | 2106 | Food Web in Florida Bay |
| | Silwood[19] | 154 | 370 | Food Web in Silwood Park |
| | Littlerock[20] | 183 | 2494 | Food Web in Little Rock lake. |
| Neuronal | C. elegans[21] | 306 | 2345 | Neural network of C. elegans |
| Transcription | E.Coli[22] | 423 | 578 | Transcriptional regulation network of Escherichia coli |
| | TRN-Yeast-1[23] | 4441 | 12873 | Transcriptional regulatory network of *S. cerevisiae* |
| | TRN-Yeast-2[24] | 688 | 1079 | Transcriptional regulatory network of *S. cerevisiae* (compiled by different group). |
| Trust | Prison inmate[25,26] | 67 | 182 | Social networks of positive sentiment |
| | Slashdot[27] | 82168 | 948464 | Social network (friend/foe) of Slashdot users |
| | WikiVote[28] | 7115 | 103689 | Who-vote-whom network of Wikipedia users |
| Electronic circuits | s208a[29] | 122 | 189 | Electronic sequential logic circuit. |
| | s420a[29] | 252 | 399 | |
| | s838a[29] | 512 | 819 | |
| Citation | ArXiv-HepTh[30] | 27770 | 352807 | Citation network of high energy physics theory in arXiv (1993-2003) |
| | SciMet[31] | 3084 | 10416 | Citation network in Scientometrics (1978-2000) |
| | Kohonen[32] | 4470 | 12731 | Citation network with topic self-organizing maps |
| WWW | Political blogs[33] | 1224 | 16718 | Hyperlinks between weblogs on US politics |
| | NotreDame[34] | 325729 | 1497134 | Web pages from University of Notre Dame |
| | BerkStan[27] | 685230 | 7600595 | Web pages from University of Notre Dame berkely.edu and stanford.edu (2002) |
| | Google[27] | 875713 | 5105039 | Web pages from Google Programming Contest |
| | Stanford[27] | 281903 | 2312497 | Web pages from Stanford University |
| Internet | p2p-1[35] | 10876 | 39994 | Gnutella peer-to-peer file sharing network (2002.08.04) |
| | p2p-2[35] | 8846 | 31839 | Gnutella peer-to-peer file sharing network (2002.08.05) |
| | p2p-3[35] | 8717 | 31525 | Gnutella peer-to-peer file sharing network (2002.08.06) |
| Organizational | Consulting[36] | 46 | 879 | Social network from a consulting company. |
| Social communication | UClonline[36] | 1899 | 20296 | Online message network of students at UC, Irvine. |
| Product co-purchasing | Amazon0302[37] | 262111 | 1234877 | Amazon product co-purchasing network (2003.0302) |
| | Amazon0312[37] | 400727 | 3200440 | Amazon product co-purchasing network (2003.0312) |
| | Amazon0505[37] | 410236 | 3356824 | Amazon product co-purchasing network (2003.0505) |
| | Amazon0601[37] | 403394 | 3387388 | Amazon product co-purchasing network (2003.0601) |
| Social network | twitter_combined[38] | 81306 | 1768149 | Social circles from Twitter (combined 973 egonets) |
| | Facebook_0[38] | 347 | 5038 | Social circles of user 0 from Facebook |

| | | | |
|---|---|---|---|
| Facebook_107[38] | 1912 | 53498 | Social circles of user 107 from Facebook |
| Facebook_348[38] | 572 | 6384 | Social circles of user 572 from Facebook |

**Table.S2**. Characteristics of the real networks analyzed in this paper. For each network, $IC_{max}$, $UMC_{max}$, $SMC_{max}$ are the relative size of the largest input, the largest unsaturated matched and the largest saturated matched control component, respectively; $p_e$ is the percentage of edges used to alter the type of the largest control component. $n_{D1}$ and $MIS_1$ are the number of possible input nodes and the size of $MIS$ before the type transition; $n_{D2}$ and $MIS_2$ are the number of possible input nodes and the size of $MIS$ after the type transition.

| Type | Name | $IC_{max}$ | $UMC_{max}$ | $SMC_{max}$ | $p_e$ | $n_{D1}$ | $n_{D2}$ | $MIS_1$ | $MIS_2$ |
|---|---|---|---|---|---|---|---|---|---|
| Food Web | StMarks | 35.19% | 38.89% | 20.37% | 3.37% | 37.04% | 100.00% | 24.07% | 3.70% |
| | Ythan | 85.19% | 5.19% | 2.96% | 10.65% | 89.63% | 4.44% | 51.11% | 3.70% |
| | Mangrove | 55.67% | 21.65% | 21.65% | 1.41% | 56.70% | 1.03% | 22.68% | 1.03% |
| | Florida Baywet | 91.41% | 6.25% | 1.56% | 1.38% | 92.19% | 0.78% | 23.44% | 0.78% |
| | Silwood | 84.42% | 1.95% | 4.55% | 29.19% | 93.51% | 9.09% | 75.32% | 5.19% |
| | Littlerock | 34.43% | 32.24% | 8.20% | 2.49% | 56.28% | 21.86% | 54.10% | 20.22% |
| Neuronal | C. elegans | 68.63% | 10.78% | 7.84% | 0.90% | 81.05% | 12.42% | 18.95% | 12.09% |
| Transcription | EColi | 0.47% | 12.53% | 9.46% | 34.78% | 73.05% | 74.94% | 72.81% | 27.19% |
| | TRN-Yeast-1 | 99.21% | 0.001% | 0.09% | 33.04% | 99.91% | 0.70% | 96.46% | 0.70% |
| | TRN-Yeast-2 | 76.60% | 0.44% | 3.63% | 41.61% | 94.91% | 18.31% | 82.12% | 16.86% |
| Trust | Prison inmate | 2.99% | 59.70% | 22.39% | 2.75% | 16.42% | 91.04% | 13.43% | 7.46% |
| | Slashdot0902 | 91.23% | 0.002% | 8.75% | 0.39% | 91.23% | 0 | 4.55% | 0 |
| | WikiVote | 0.03% | 32.12% | 1.25% | 3.60% | 66.59% | 93.37% | 66.56% | 14.17% |
| Electronic circuits | s208a | 17.21% | 7.38% | 9.84% | 5.82% | 33.61% | 16.39% | 23.77% | 14.75% |
| | s420a | 9.13% | 4.37% | 4.76% | 3.26% | 32.94% | 23.81% | 23.41% | 18.25% |
| | s838a | 5.27% | 2.15% | 2.34% | 2.08% | 32.62% | 27.34% | 23.24% | 19.92% |
| Citation | ArXiv-HepTh | 14.44% | 48.96% | 16.71% | 0.91% | 33.98% | 85.87% | 21.58% | 10.06% |
| | SciMet | 52.14% | 6.55% | 16.28% | 7.62% | 64.40% | 12.26% | 37.48% | 11.74% |
| | Kohonen | 60.25% | 0.27% | 13.62% | 14.45% | 66.91% | 6.67% | 47.29% | 6.13% |
| WWW | Political blogs | 52.61% | 19.53% | 13.40% | 1.46% | 66.91% | 14.30% | 34.15% | 14.22% |
| | NotreDame | 53.90% | 1.16% | 6.01% | 9.15% | 87.04% | 33.14% | 67.71% | 25.64% |
| | BerkStan | 57.27% | 7.77% | 13.99% | 2.78% | 73.21% | 15.94% | 65.69% | 34.83% |
| | Google | 60.80% | 8.84% | 11.68% | 3.97% | 73.58% | 12.78% | 36.95% | 13.83% |
| | Stanford | 63.50% | 3.56% | 19.36% | 2.83% | 72.13% | 8.63% | 35.91% | 12.66% |
| Internet | p2p-1 | 90.58% | 1.34% | 7.13% | 14.88% | 91.11% | 0.52% | 55.20% | 0.47% |
|  | p2p-2 | 90.55% | 0.97% | 6.15% | 15.52% | 92.64% | 2.09% | 57.78% | 1.92% |
|  | p2p-3 | 91.75% | 0.76% | 5.79% | 15.58% | 93.25% | 1.50% | 57.74% | 1.40% |
| **Organizational** | Consulting | 97.83% | 0 | 2.17% | 0.23% | 97.83% | 0 | 4.35% | 0 |
| **Social communication** | UClonline | 79.94% | 3.84% | 12.95% | 2.84% | 81.89% | 1.95% | 32.33% | 1.95% |
| **Product co-purchasing networks** | Amazon0302 | 8.76% | 49.55% | 32.32% | 0.30% | 17.74% | 92.32% | 3.23% | 1.79% |
|  | Amazon0312 | 0.95% | 3.04% | 83.61% | 0.00003% | 12.71% | 88.45% | 3.52% | 3.52% |
|  | Amazon0505 | 91.35% | 1.72% | 6.08% | 0.44% | 91.46% | 0.11% | 3.62% | 0.05% |
|  | Amazon0601 | 0.10% | 75.90% | 18.74% | 0.21% | 5.29% | 96.02% | 2.04% | 0.27% |
| **Social network** | twitter_combined | 79.40% | 7.23% | 12.37% | 0.88% | 80.04% | 0.63% | 19.39% | 0.34% |
|  | Facebook_0 | 0.86% | 0.58% | 86.46% | 0.02% | 7.49% | 89.34% | 5.48% | 5.48% |
|  | Facebook_107 | 0.05% | 0 | 54.08% | 0.002% | 45.92% | 99.16% | 45.92% | 45.92% |
|  | Facebook_348 | 0.35% | 0.17% | 38.64% | 0.02% | 61.19% | 99.48% | 61.01% | 61.01% |